\documentclass{ws-rv975x65}
\usepackage{subfigure}     
\usepackage[square]{ws-rv-van}     
\makeindex
\usepackage{amssymb,amsmath,amsfonts,dsfont}

\begin{document}

\chapter[Seeing spin dynamics in atomic gases]{Seeing spin dynamics in atomic gases}\label{ch:dmsk}

\author[Dan M. Stamper-Kurn]{Dan M. Stamper-Kurn}

\address{Department of Physics, University of California, Berkeley, California 94720, USA \\
Materials Sciences Division, Lawrence Berkeley National Laboratory, Berkeley, California 94720, USA}

\begin{abstract}
The dynamics of internal spin, electronic orbital, and nuclear motion states of atoms and molecules have preoccupied the atomic and molecular physics community for decades.  Increasingly, such dynamics are being examined within many-body systems composed of atomic and molecular gases.  Our findings sometimes bear close relation to phenomena observed in condensed-matter systems, while on other occasions they represent truly new areas of investigation.  I discuss several examples of spin dynamics that occur within spinor Bose-Einstein gases, highlighting the advantages of spin-sensitive imaging for understanding and utilizing such dynamics.
\end{abstract}

\body

\section{Atomic materials science}\label{sec:intro}

It is conventional to begin a book chapter such as this with a coherent framing of the progress within a scientific field, perhaps portrayed as setting up the next clear question that must be answered or an opportunity that must be exploited.  However, it has always struck me that such a presentation belies the true and exciting nature of scientific work, in which discoveries arise from unframed curiosity and in which the intellectual random walks of many individuals cause the bounds of our knowledge and our sense of what is worth knowing to diffuse outward.  Read as a whole, the chapters of this book support this view of science as a diffusing endeavor.  One may try to tie all the work described herein as a coherent body of work (about coherence).  But maybe it is best to appreciate the diverse manners in which the field of AMO science is evolving, all the fun that people are having in their laboratories and offices, and all the places we may go in the coming years.

This Chapter is about spin dynamics in quantum degenerate gases.  I shall emphasize the role of high-resolution direct imaging to visualize such dynamics.  Direct imaging provides a great deal of detailed information about the gases being probed.  While we have learned a few ways in which some portion of that information can be analyzed, there still remains a wealth of information to be mined from detailed images.  Above all, however, the direct images of spin dynamics have the value of being beautiful, of portraying the science at hand as an appealing subject to study.

Now, to keep with convention, let me reframe the study of spin dynamics in quantum gases in a coherent framework.

Simple, discrete quantum systems, along with methods that make use of resonance and coherence to manipulate and measure such systems, have been the bread and butter of atomic physics for a century.  In experiments on atomic beams and vapor cells, and then certainly on trapped ions, the focus has often been on the internal dynamics within single atoms.  Precise measurements of these dynamics have allowed us to test the principles by which atoms are constructed (e.g.\ quantifying high-order processes in quantum electrodynamics), or to characterize the conditions to which the atoms are exposed (e.g.\ sensing magnetic fields).

Increasingly, however, atomic physics is giving access to many-body systems in which the atoms evolve no longer in isolation, but rather as a collection of mutually interacting quantum objects.  In other words, we are moving from the study of atoms to the study of materials.  The focus of materials science is typically different than that of atomic physics: whereas the atomic physicist may appreciate a precise characterization of the full dynamics of a simple quantum system, the material scientist may focus instead on general characterizations of thermodynamic ground states and of small excitations atop such equilibria.  Precise characterization of materials is often unwarranted since material samples are imperfect and differ from one another.  Characterizing far-from-equilibrium states is unnecessary since most materials cannot typically be driven very far from equilibrium.

Now that atomic physicists are creating materials-like quantum systems under highly controlled conditions, both scientific approaches are beneficial and applicable.  It is believed that atomic systems, serving as ``quantum emulators'' of paradigmatic many-body quantum systems, making use of the tremendous control, cleanliness, and powerful detection capabilities offered by atomic physics, may allow us to determine better the principles underlying the materials science of both existing and theorized matter.  Conversely, one expects that the properties of complex materials realized by these atomic systems will make it possible for us to improve our performance in the typical tasks of an atomic physicist, vis a vis precision measurement and sensing.

This duality is exhibited by the study of magnetism in quantum gases, which is the topic of this Chapter.  In particular, I focus on magnetic phenomena in spinor Bose-Einstein gases\index{spinor Bose-Einstein gas!definition}, ones composed of bosonic atoms that are free to occupy the various Zeeman sublevels of a manifold of spin states.  Following the example of materials science, I describe the propagation of small-amplitude spin modulations in a ferromagnetic condensate (Sec.\ \ref{sec:magnons}) in terms of the dispersion of magnon excitations, and the spin dynamics of non-equilibrium quantum fluids (Sec.\ \ref{sec:spinmixing}) in terms of spontaneous symmetry breaking and phase transitions.  Following the example of atomic physics, I describe the propagation of magnons (Sec.\ \ref{sec:magnons}) as a promising form of atomic interferometry, the spin-mixing instability (Sec.\ \ref{sec:spinmixing}) in terms of the generation of spin-nematicity squeezed states, and the combination of Larmor precession and high-resolution, high-sensitivity imaging (Secs.\ \ref{sec:imaging}, \ref{sec:magnetometry}) as a potentially powerful magnetic field sensor.

This Chapter is not meant to be a comprehensive review of studies of magnetic phenomena in quantum gases, or even specifically of research on spinor Bose-Einstein gases.  Such reviews are provided elsewhere (e.g.\ in Refs.\ \refcite{bloc08rmp,bloc12qsim,stam13rmp}).  Rather, the intention is to highlight some of the new research directions in materials science and atomic physics that have been identified in recent years and that should, I believe, be taken up (by the reader!) in the coming years.

\section{Imaging methods}
\label{sec:imaging}

\index{imaging methods!time of flight|(}One of the standard ways to diagnose an ultracold gas is to let it out of its trapping container, allow it to expand a while, and then to take a picture that reveals the spatial distribution of the expanded cloud.  This technique is favored because the expanded gas can be made far larger than the imaging resolution, so as to avoid imaging aberrations.  Also, the expanded gas can be made to have a low optical density.  One can illuminate the sample with light that is resonant with an atomic transition, and then quantify the resulting shadow image to determine faithfully the column density distribution.  This distribution carries information on an amalgam of properties of the gas -- such as the one-particle distribution and two-particle correlations in position and momentum space, the interaction energy, angular momentum, vorticity, etc.\ -- depending on the time of flight.  That is, the spatial distribution at zero time of flight is obviously the in-trap spatial distribution, but understanding which property is revealed by which feature in the time-of-flight distribution for longer expansion times depends on assumptions about the behavior of the gas upon expansion.  For gases that are structured and correlated with greater complexity, the connection between time-of-flight spatial distributions and in-situ properties is less straightforward.

The time-of-flight imaging method can be made spin sensitive by applying a magnetic field gradient during the time of free expansion, emulating Stern and Gerlach's early experiment in which atoms in different Zeeman states were deflected to different spatial regions on a detector \cite{gerl24}.  Separating the different Zeeman components from one another requires a non-zero time of flight, so that, aside from the overall population distribution among the Zeeman sublevels, all other information about the gas is ambiguous.  In particular, the spatial spin distribution can be determined only with poor spatial resolution.\index{imaging methods!time of flight|)}

My group has developed imaging methods that allow for the spin distribution of a quantum gas to be measured directly, in-situ, with high spatial resolution and high spin sensitivity.  Both methods make use of the atomic physics of single atoms to characterize the magnetic structure of a many-body system: The first method relies on optical birefringence, while the second relies on the spin-selectivity of microwave transitions between different hyperfine states within the electronic ground state.

\subsection{Spin-sensitive dispersive in-situ imaging}

\index{imaging methods!spin-sensitive dispersive|(}Birefringent materials are ones for which the real optical susceptibility is a tensor, rather than a scalar; in other words, light passing through the material acquires a phase shift that depends on the polarization of the light, and, for this reason, may emerge from the material with a different polarization than it had before entering the material.

One distinguishes between linear and circular birefringence.  In the case of linear birefringence, if we represent the polarization of light in a particular basis of linear polarizations, the optical susceptibility is a diagonal matrix, meaning that light entering the material with either of these linear polarizations acquires a polarization-dependent phase shift but does not change in polarization.  For example, the birefringent materials used in optical waveplates are linearly birefringent, defined by orthogonal ``fast'' and ``slow'' axes.  In the case of circular birefringence, the basis polarizations that diagonalize the susceptibility tensor are the left- and right-handed circular polarizations.  Circular birefringence occurs in situations that break time-reversal symmetry, e.g.\ in optical isolators that use strong magnetic fields and the Faraday effect.

\index{atomic optical birefringence|(}A spin-polarized atomic gas can exhibit both linear and circular birefringence.  For example, consider a gas fully polarized in the $|m_J = 1/2\rangle$ sublevel (with respect to some spatial quantization axis) of a $J = 1/2$ ground state.  The oscillator strength for an optical transition to either a $J^\prime = 1/2$ or $J^\prime = 3/2$ excited state differs for $\sigma^+$ and $\sigma^-$ polarized light; this can be seen by examining the values of Clebsch-Gordan coefficients.  Thus, the dispersive phase shift imposed on light that passes through such an atomic gas, and that is detuned from the optical transition, is different for the two circular optical polarizations.  This difference quantifies the spin polarization of the atomic gas.  As another example, consider a gas fully polarized in the $|m_J = 0\rangle$ sublevel of a $J=1$ ground state, and probed near a transition to a $J^\prime = 1$ excited state.  This gas does not couple to $\pi$-polarized light (linear along the quantization axis), but does couple to the orthogonal linear polarization.  Thus, such a gas exhibits linear birefringence which, in this case, identifies the spin quantization axis.

It can be shown that circular birefringence reveals the spin-vector moments while linear birefringence reveals the spin-quadrupole moments of the atoms being probed \cite{sute05book}\index{spinor Bose-Einstein gas!magnetization}\index{spinor Bose-Einstein gas!nematicity}.  These moments yield the magnetization vector and nematicity tensor of the material comprised of such atoms.  The spin-vector moments are associated with coherences between Zeeman sublevels that differ by $\Delta m = 1$ in their magnetic quantum number, while the spin-quadrupole moments are associated with $\Delta m = 2$ coherences.  That is why I could use the example of a $J=1/2$ ground state to discuss circular birefringence, but needed a $J=1$ ground state, which can have $\Delta m=2$ coherences, to exemplify linear birefringence.

\index{atomic optical birefringence|)}

\begin{figure}[t]
\begin{center}
\includegraphics[width=4.5in]{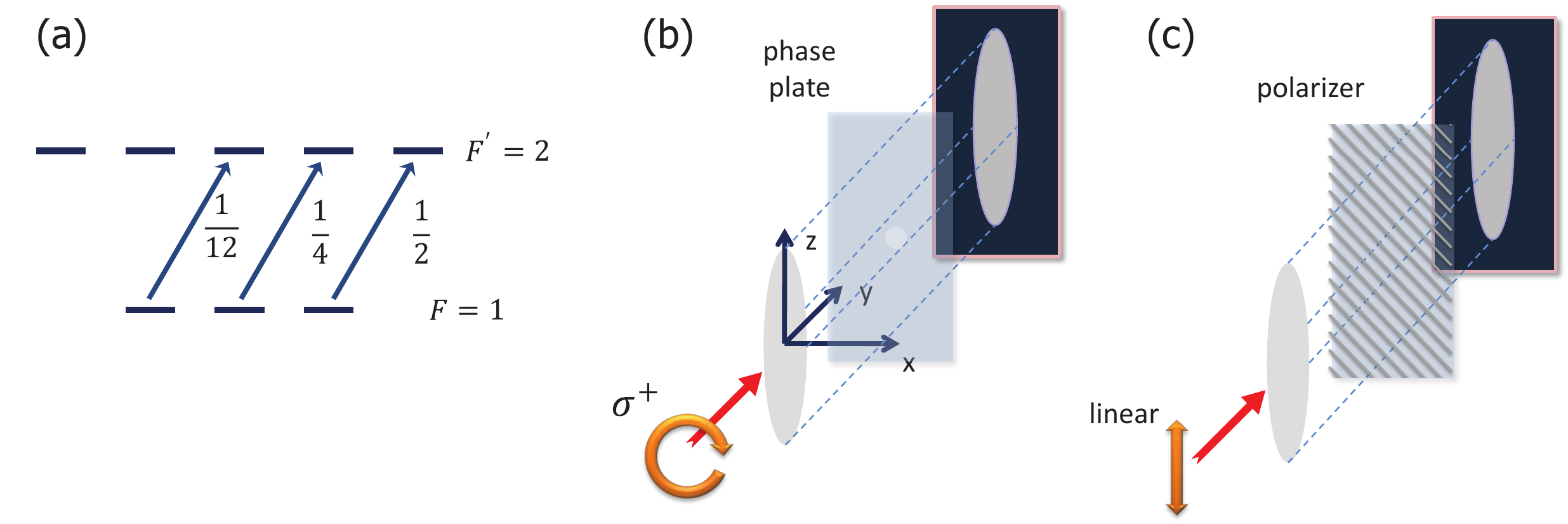}
\end{center}
\caption{We employ circular birefringence to image the magnetization of an $F=1$ spinor gas.  (a) The relative oscillator strengths for circular polarized light driving an optical transition from the $F=1$ ground state to an $F^\prime = 2$ excited state are shown.  Atoms polarized in the $|m_F=+1\rangle$ sublevel, quantized along the direction of the optical helicity, interact more strongly with the light.  A dispersive image taken with light near this transition therefore contains information on the magnetization of the gas along that direction.   The dispersive signal is turned into an image in one of two standard ways.  (b) One way is to convert the phase shift on a circularly polarized light field into an intensity image using a phase plate placed in the Fourier plane of the image, this being a form of phase contrast imaging \cite{hech89}.  (c) A second way is to illuminate the sample with linear polarized light.  The circular birefringence causes a rotation of the linear polarization, and this rotation is analyzed by passing the light through a linear polarizer before the camera.}
\label{fig:birefimaging}
\end{figure}

\begin{figure}[t]
\begin{center}
\includegraphics[width=4in]{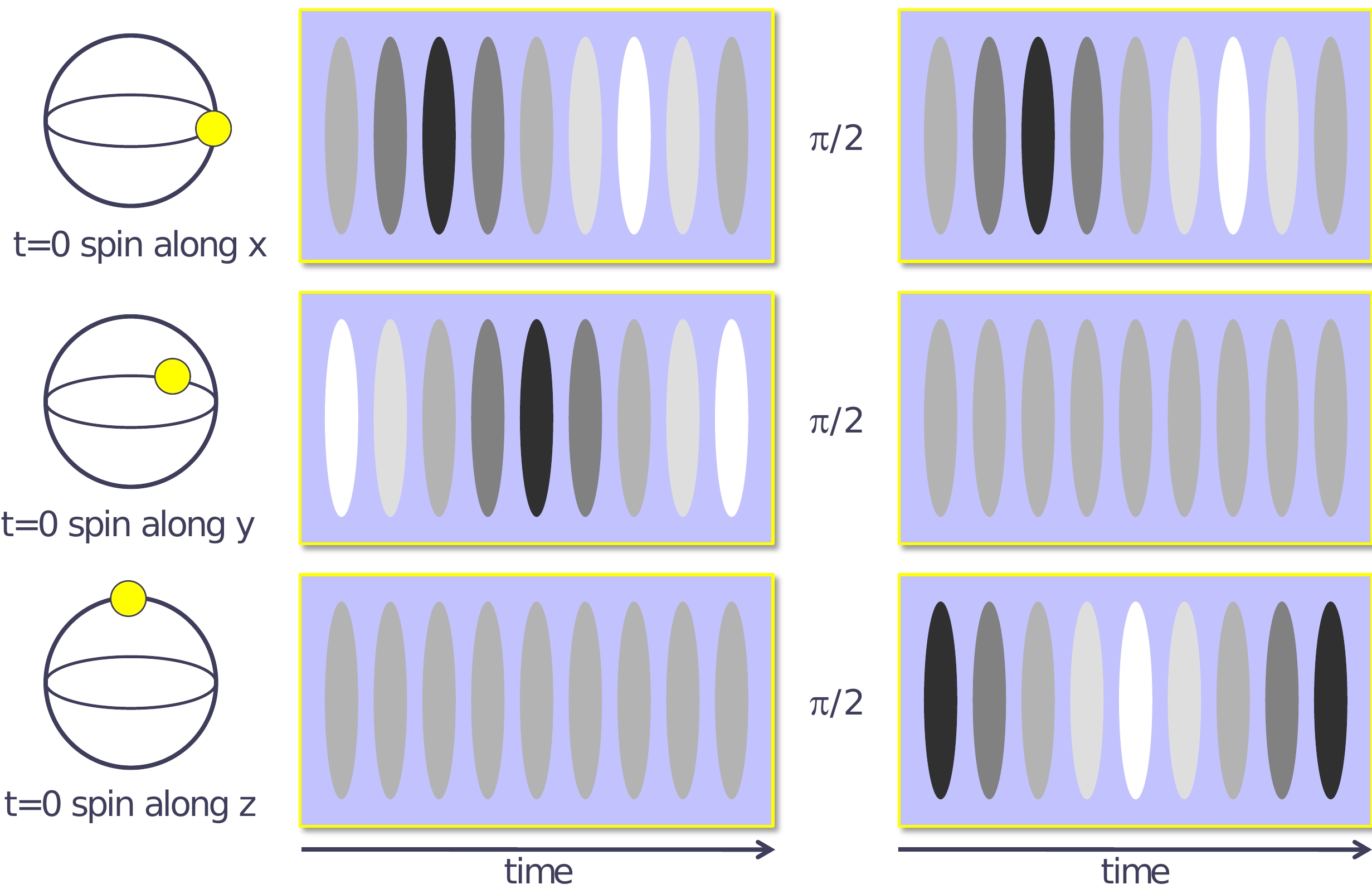}
\end{center}
\caption{All three components of the vector magnetization can be quantified by taking a sequence of imaging frames while the gas magnetization precesses in an applied magnetic field, and then is re-oriented using rf pulses.  Shown is the expected image frame sequence for a gas initially magnetized along each of the cardinal directions.  The first set of panels shows the expected signal during nine equally spaced image frames taken during Larmor precession.  The image signal from transversely magnetized sample shows a temporal oscillation, the phase of which determines the direction of the magnetization within the transverse plane.  Following a $\pi/2$ rf pulse, which causes the longitudinal magnetization to be rotated into the transverse plane, a second set of several image frames can be analyzed to quantify that transverse magnetization strength.  A two-dimensional image of the vector magnetization is reconstructed from such an imaging sequence.  The entire sequence occurs within just a few ms, faster than the dynamical timescales for spin transport and mixing dynamics.}
\label{fig:lpimaging}
\end{figure}

My group made use of the circular birefringence of $^{87}$Rb atoms in the $F=1$ hyperfine ground state in order to image their magnetization\cite{higb05larmor}.  Polarized light detuned by several hundred MHz from the D1 optical transitions was sent along the thin axis of a scalene ellipsoidal optical trap containing a gas of several million atoms, which were typically deep within the quantum degenerate regime (Fig.\ \ref{fig:birefimaging}).  In one approach, we imaged with a brief pulse of circular polarized light, and used phase-contrast imaging to convert the spatially varying phase shift on the probe light into an intensity image.  A single image yielded information on both the density and the column-integrated magnetization of the gas along the probe axis.  To isolate the magnetization signal, and also characterize the projections of the magnetization along all axes, we used a combination of rf pulses and Larmor precession to rotate the magnetization (Fig.\ \ref{fig:lpimaging}).  In a second approach, we used linearly polarized light and quantified its polarization rotation using a linear polarizer before the camera.  The image now measured only the magnetization.  Again, a sequence of rf-pulses was used to bring each of the magnetization components into view \cite{guzm11}.

\renewcommand{\baselinestretch}{1}
\renewcommand{\baselinestretch}{1.5}

\renewcommand{\baselinestretch}{1}
\renewcommand{\baselinestretch}{1.5}

\begin{figure}[t]
\begin{center}
\includegraphics[width=2.5in]{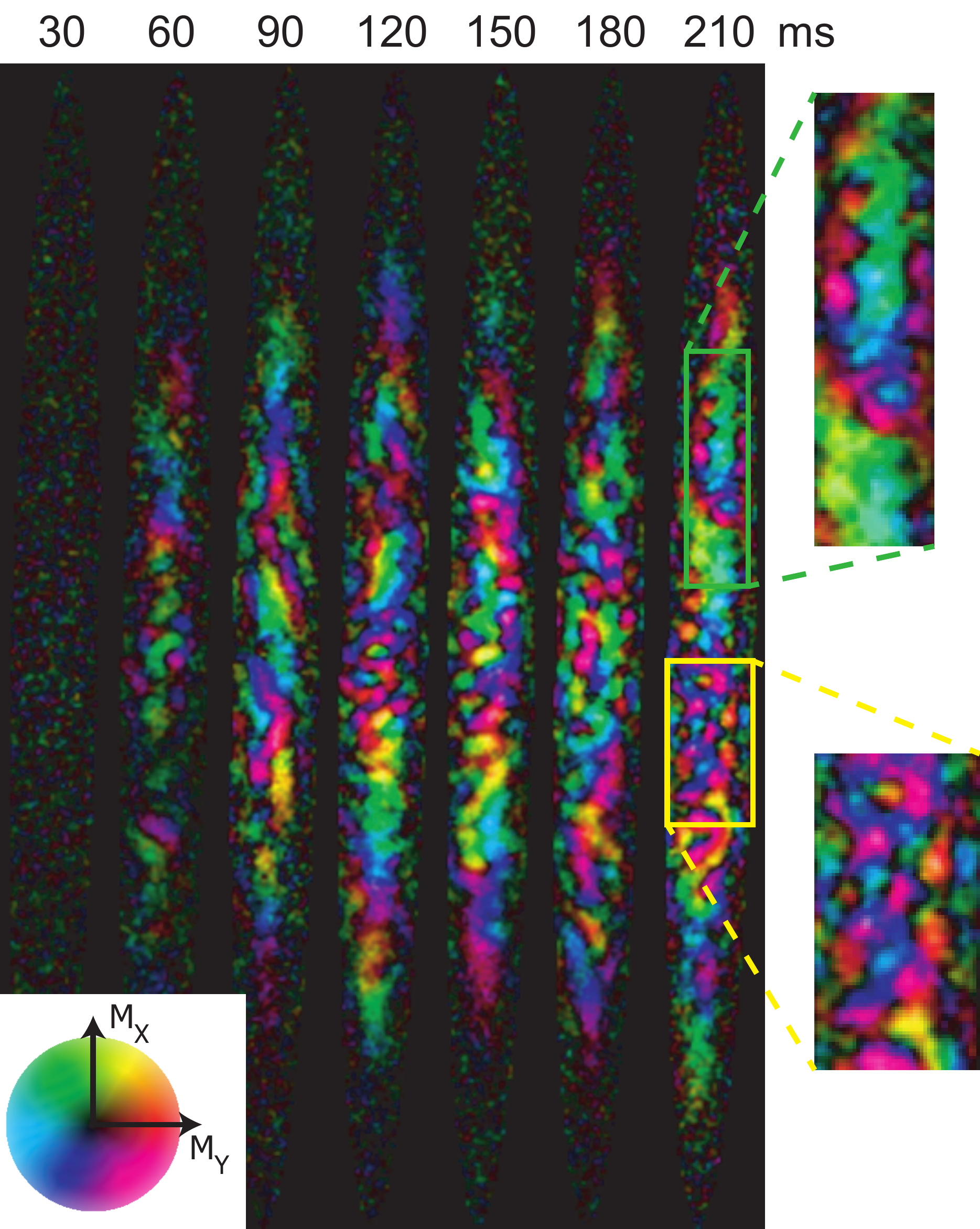}
\end{center}
\caption{The transverse magnetization of a spinor Bose-Einstein condensate was measured at a variable time (indicated at top of image) after it was quenched across a paramagnet-to-ferromagnet phase transition.  The quantum degenerate gas was initially allowed to equilibrate in the $|m_F = 0\rangle$ state at a high quadratic Zeeman shift $q$ where that state is the equilibrium spin state.  The quench was performed by rapidly lowering the quadratic Zeeman energy to near zero, encouraging the formation of a transversely magnetized ferromagnetic condensate.  The data are encoded in a color scheme (see color wheel at lower left) where the transverse magnetization orientation is indicated by the hue, and its magnitude by the color saturation.  The image of each separate condensate has a width of about 30 microns.  The magnetization appears over 10's of ms after the quench, resulting in the formation of some regions with small neighboring spin domains of opposite orientation (highlighted by the yellow box and magnified for view for the latest time image) and also large domains of common orientation (green box).  The figure is adapted from Ref.\ \refcite{sadl06symm}.}
\label{fig:quench}
\end{figure}

The quality of data obtained by this method is exemplified by Fig.\ \ref{fig:quench}, which reveals the response of a quantum degenerate gas to being quenched across a paramagnet-to-ferromagnet phase transition \cite{sadl06symm}\index{spinor Bose-Einstein gas!phase transition}\index{quantum quench}.  The physics revealed by these images is discussed below in Sec.\ \ref{sec:spinmixing}.  Here, let me just emphasize how eye-opening these images were.  Prior studies of spin-mixing dynamics in rubidium spinor gases \cite{chan04,schm04}, relying on Stern-Gerlach time-of-flight imaging, had revealed damped oscillations in the Zeeman-state populations, with very little spatial resolution.  Theorists guessed that the damping was due to some sort of multi-mode spatial dynamics \cite{murp06finiteT}, but it was impossible to tell from the data whether that was the proper explanation.  In any case, such multi-mode effects did not seem like something interesting to examine further, but rather were seen as something to try to eliminate by working with smaller and more tightly confined samples\footnote{The strategy of restricting the sample geometry to achieve strictly single-mode dynamics has been pursued and has turned out to be very beneficial for isolating and controlling the many-body-quantum nature of spin-mixing dynamics; see Refs.\ \refcite{haml12squeezing,luck14entangle,stro14fisher} for some examples.}. In contrast, our high-resolution imaging of spin distributions revealed a rich dynamics: the development of spin domains with magnetization lying in all directions transverse to the magnetic field direction, separated by snaking domain walls and pierced by spin vortices.

\renewcommand{\baselinestretch}{1}
\renewcommand{\baselinestretch}{1.5}

\index{imaging methods!spin-sensitive dispersive|)}Indeed, the progress of ultracold atomic physics over the past couple of decades has shown, time and time again, that improved imaging methods allow us to examine interesting physical phenomena whose study was previously inaccessible, and, indeed, inconceivable.  Examples of such leaps include the first realization of non-destructive imaging \cite{andr96} that allowed for studies of bosonic stimulation in the formation of a Bose-Einstein condensate \cite{mies98form}, high-resolution objectives that allowed for the realization of Josephson junctions formed by optical potentials \cite{levy07josephson} and observations of scale invariance \cite{hung11invariance} and quantum criticality \cite{zhan12qcrit} in two-dimensional gases, and the recent studies of dynamics and magnetism in the single-site-resolving quantum gas microscope \cite{bakr09microscope,sher10singleatom}.  I am certain this trend will continue.

\subsection{Absorptive spin-sensitive in-situ imaging}

\index{imaging methods!absorptive spin-sensitive in-situ imaging|(}Dispersive imaging has been billed as a form of ``non-destructive imaging,'' implying that imaging the gas does not deposit so much energy so as to utterly destroy it.  Perhaps a better moniker is ``less-destructive imaging,'' for the act of imaging necessarily disturbs the gas.  There is a tradeoff between how much information one wants to extract the gas and how much one wants to preserve the gas for continued evolution and imaging.  The off-resonant probe light used in dispersive imaging perturbs the gas in several ways.  Characterizing the vector magnetization, and even more so the tensor nematicity, of a gas requires a sequence of several images, and, thus, that each image be non-destructive enough to leave enough signal for the subsequent image.  Therefore, there is an upper bound on the signal to noise with which these characteristics can be quantified.  These bounds were explored in my group's work on spatially resolved magnetic field sensing, which is discussed further in Sec.\ \ref{sec:magnetometry}.

\begin{figure}[t]
\begin{center}
\includegraphics[width=5.5in]{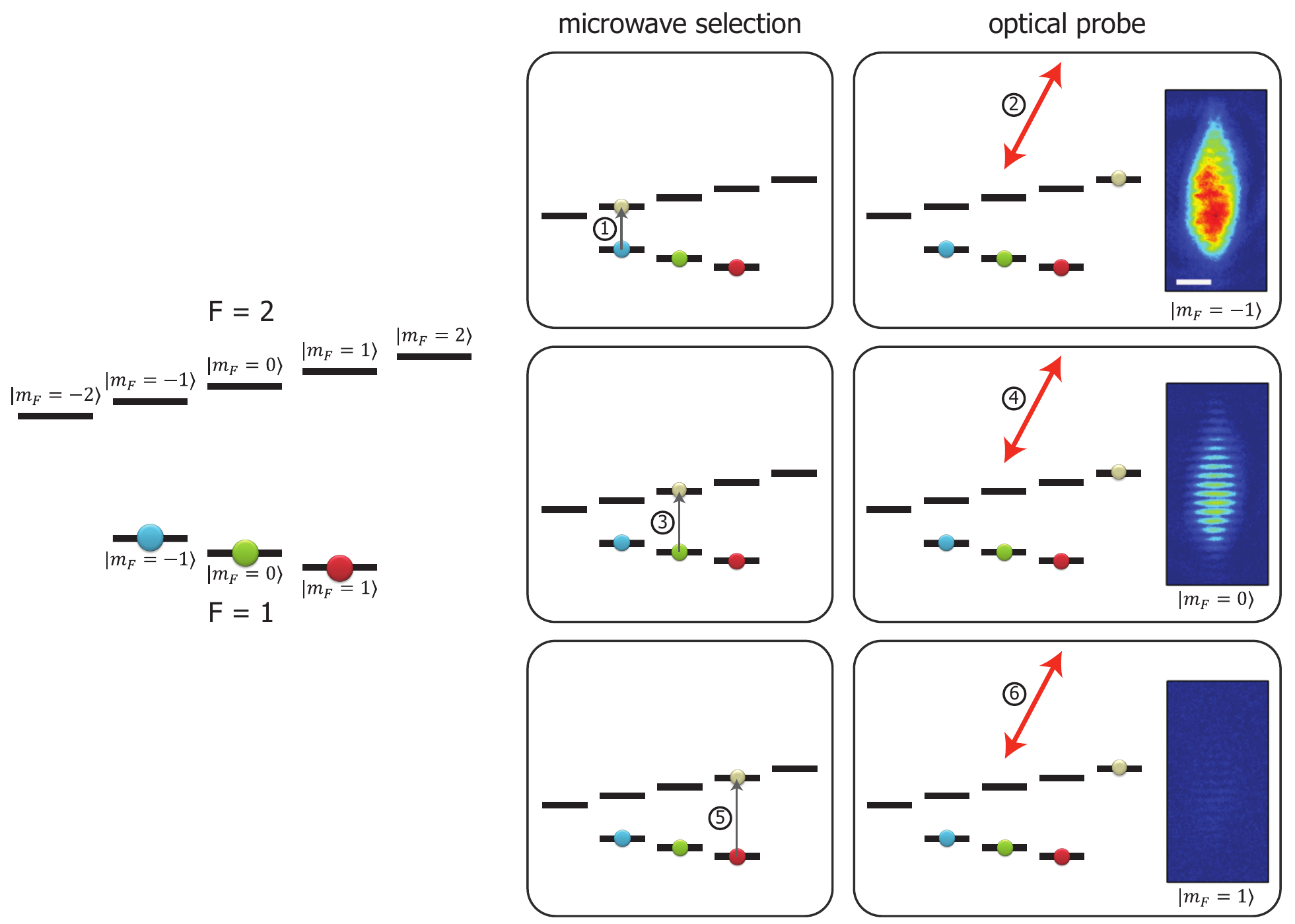}
\end{center}
\caption{Absorptive spin-selective in-situ imaging of an $F=1$ rubidium gas.  Left: The ground electronic state has two hyperfine states, denoted by quantum numbers $F=1$ and $F=2$.  Here the Zeeman sublevels within each manifold are separated by the linear Zeeman shift, which is shown out of scale to the very large energy separation between hyperfine spin manifolds.  Our task is to image a gas that occupies the $F=1$ levels.  This is accomplished in six steps.  First, a short microwave pulse is used to drive a small fraction of atoms from one of the $F=1$ magnetic sublevels to the unoccupied $F=2$ state.  Second, an optical probe measures the in-situ column density of the selected atoms.  In the process the atoms are optically pumped, so that the absorption image is quantitative, and also the atoms are heated sufficiently so that they are expelled from the trap after imaging.  This selection/probe sequence is repeated three times to gather three image frames.  The images show a standing wave of magnons in a ferromagnetic initial state.  The scale bar indicates a 50 $\mu$m distance.}
\label{fig:assisi}
\end{figure}

The limitations we identified led us to consider an alternative imaging scheme: absorptive spin-sensitive in-situ imaging (ASSISI).  The idea here is to analyze the population of atoms in one set of atomic levels $|A\rangle$ by promoting a small fraction of them to unoccupied atomic levels $|B\rangle$, and then to image those promoted atoms selectively with probe light that is resonant with levels $|B\rangle$ but far detuned from levels $|A\rangle$.  The promoted atoms are allowed to scatter very many photons, allowing one to obtain a low-noise image of their distribution.  After the image is taken, the imaged atoms are expelled from the trap, leaving the remaining sample of atoms relatively unperturbed.  The application of this scheme to imaging the $|F=1\rangle$ hyperfine states of $^{87}$Rb is described in Fig.\ \ref{fig:assisi}.  The signal-to-noise ratio of this imaging method is found to be limited by shot-noise fluctuations in the number of promoted atoms, and vastly improved over the quality of dispersive imaging.

\renewcommand{\baselinestretch}{1}
\renewcommand{\baselinestretch}{1.5}

Another important benefit of ASSISI is that one can fully characterize the atomic spin state.  In particular, the dispersive imaging method described previously permits measurements of the spin-quadrupole moment only through the linear birefringence of the atomic gas.  For alkali atoms, this linear birefringence is weak except for light very close to atomic resonances, where the absorption of the probe light causes severe problems.  In contrast, ASSISI permits a complete in-situ ``Stern-Gerlach'' analysis of the Zeeman sublevels along any axis.  Performing such analyses along several axes permits one to determine the atomic spin state completely, including the $\Delta m=2$ coherences associated with the spin-quadrupole moments, and even higher-order coherences in the case of high-spin gases.\index{imaging methods!absorptive spin-sensitive in-situ imaging|)}

\section{Coherent propagation of magnons}
\label{sec:magnons}

\index{magnons|(}
Now I turn to the first of three examples where high-resolution in-situ imaging has been useful for elucidating magnetic phenomena in quantum gases, and where such phenomena have been examined both from the viewpoint of materials science and of atomic physics.  We start with understanding how magnetic excitations propagate in such gases.

\index{spinor Bose-Einstein gas!interactions}\index{spinor Bose-Einstein gas!ground state}Ultracold spinor Bose-Einstein gases will condense into a state that is both magnetically ordered and also superfluid.  After all, it is the magical nature of Bose-Einstein condensation to cause particles to occupy the lowest-energy single particle states with a macroscopic occupancy, even at temperatures $T$ where the energy difference between the lowest and next-lowest energy is much smaller than the thermal energy available $k_B T$.
Given this tendency, one expects the spinor Bose-Einstein condensates to adopt the single-particle spin state with the lowest energy.  Absent any external spin-dependent influences such as applied magnetic fields, the dominant spin-dependent energy that remains is the contact interaction, which describes the effect of the short-range interaction potential between the neutral atoms colliding at very low incident energy.  This interaction can be expressed by the operator $C(\mathbf{F}_i, \mathbf{F}_j) \delta^3(\mathbf{r}_i - \mathbf{r}_j)$ where $i$ and $j$ are the indexes of the two atoms colliding, $\mathbf{F}_{i,j}$ are their spin operators and $\mathbf{r}_{i,j}$ their positions.  We know that $C(\mathbf{F}_i, \mathbf{F}_j)$ must be rotationally symmetric\index{spinor Bose-Einstein gas!symmetry}.  Here I have neglected the magnetic dipole interactions, which are treated as a long-range interaction and are rotationally symmetric only under rotations both in position and spin space.\index{spinor Bose-Einstein gas!magnetic dipolar interactions}

To keep the discussion focused, I will concentrate on the specific example of the $F=1$ spinor gas of $^{87}$Rb, for which one finds $C(\mathbf{F}_i, \mathbf{F}_j) = c_0^{(1)} + c_1^{(1)} \mathbf{F}_1 \cdot \mathbf{F}_2$, with $c_1^{(1)}<0$\footnote{The quantity $c_1^{(1)}$ is determined by differences in s-wave scattering lengths for collisions in different angular momentum channels, which are in turn determined by the particular form of the atom-atom interaction potential, which is in turn determined by comparison to a variety of measurements.}.  Such an interaction favors ferromagnetic ordering, in which all atoms' spins are commonly oriented so as to minimize the contact interaction energy.  For this reason, we may denote the $F=1$ $^{87}$Rb quantum gas as a ferromagnetic superfluid.

\subsection{Nambu-Goldstone bosons of magnetically ordered superfluids}
\label{sec:nambugold}

\begin{figure}[t]
\begin{center}
\includegraphics[width=5in]{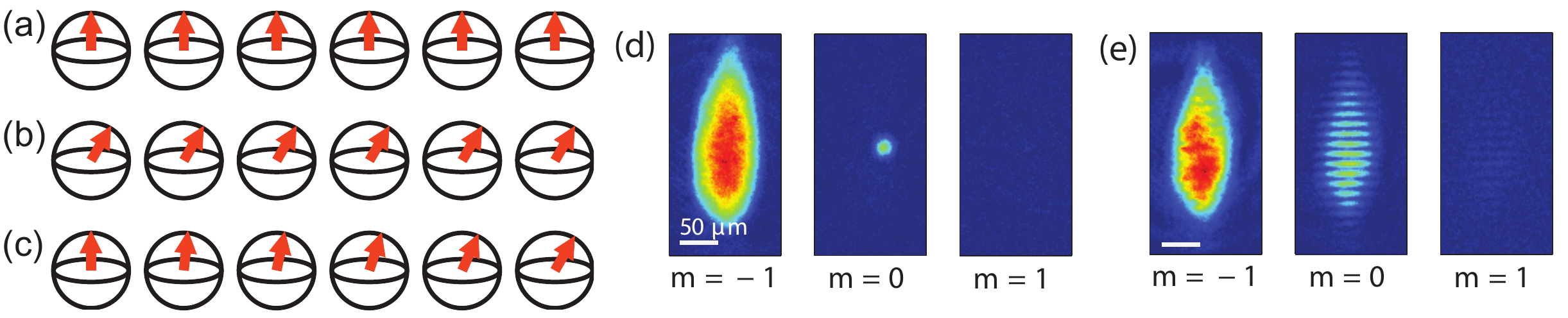}
\end{center}
\caption{Magnon schematic (left): (a) A material equilibrates in a uniform broken symmetry state (e.g.\ a ferromagnet with magnetization pointing up).  (b) Applying a uniform symmetry transformation (e.g.\ rotating the magnetization) does not change the free energy.  (c) This suggests that an inhomogeneous transformation of the system, which varies only over long length scales, costs little extra energy.  This excitation is the Nambu-Goldstone boson, which, in the absence of long-range interactions, is gapless at the limit of zero wavevector.  Magnon images (right): Zeeman-state populations are characterized in-situ by a sequence of microwave excitation and resonant optical absorption imaging.  (d) A localized coherent wavepacket is created using the vector ac Stark shift of light focused tightly within the center of the condensate.  (e) A spatial grating of magnons with well defined wavevectors is created using two beams of light that form a high-contrast interference pattern at the location of the trapped gas.  For weak magnon excitations atop a longitudinally polarized gas, the $m_F=0$ population is representative of the magnon population.}
\label{fig:magnonimage}
\end{figure}

\index{Nambu-Goldstone bosons|(}Systems that spontaneously break a continuous symmetry commonly possess gapless bosonic collective excitations, known as Nambu-Goldstone bosons.  To visualize such excitations, we consider an equilibrium state in which a continuous symmetry $S$ is broken; e.g.\ a ferromagnet with uniform magnetization  (Fig.\ \ref{fig:magnonimage}a).  If one applies an infinitesimal operation within $S$ throughout the system, the energy is unchanged.  Now consider that one transforms the system by applying an infinitesimal symmetry operation that varies in space with a well-defined wavevector $\mathbf{k}$.  Such a transformation defines a bosonic collective excitation: the Nambu-Goldstone boson.  This excitation may be expected to raise the energy of the system, since now the order parameter is nonuniform.  However, in the limit $\mathbf{k} \rightarrow 0$, the transformed system is nearly uniform, so the rise in energy should tend to zero, implying the boson is gapless.  The Nambu-Goldstone theorems do not hold for systems with long-range interactions, e.g.\ for charged superfluids and for the broken electroweak symmetry.

\renewcommand{\baselinestretch}{1}
\renewcommand{\baselinestretch}{1.5}

In recent years, the Nambu-Goldstone theorems have been re-examined in an attempt to reconcile the phenomena of symmetry breaking in relativistic and non-relativistic scenarios:  In particle physics, excepting the Higgs mechanism, each broken symmetry leads to a gapless linearly dispersing bosonic mode.  In contrast, in material systems, not every broken symmetry produces a new boson, and many of these bosons disperse quadratically, rather than linearly.  The mathematical conditions leading to these differences have only recently been identified \cite{wata12unified,wata13redund}.

Now we return to considering the ferromagnetic superfluid realized by the $F=1$ $^{87}$Rb gas. It breaks three symmetries: a combination of gauge symmetry and rotational symmetry, and the symmetries of rotations about the transverse $\mathbf{x}$ and $\mathbf{y}$ axes.  \index{spinor Bose-Einstein gas!symmetry}  While one might naively expect the ferromagnetic state to carry three Nambu-Goldstone bosons, in fact there are just two.  The breaking of gauge/rotation symmetry (rotation about $\mathbf{z}$ gives a phase shift) is the origin of the gapless, linearly dispersing phonon.  The consequence of the breaking of transverse rotational symmetries is different.  Because the generators of the two broken symmetries have a commutator with a non-zero expectation value in the ground state \cite{wata12unified}, the two symmetries generate only a single Nambu-Goldstone boson, and the dispersion relation of this boson is quadratic with wavevector, i.e.\ $E(\mathbf{k}) = \hbar^2 k^2 / (2 m^*)$\index{magnons!effective mass}\index{magnons!dispersion relation}.  This boson is the ferromagnetic magnon.  It corresponds to rotating the uniform ground state by an infinitesimal angle $\theta$ about the transverse axis $\mathbf{n}_\phi = \mathbf{x} \cos\phi + \mathbf{y} \sin\phi$ with $\phi = \mathbf{k} \cdot \mathbf{r}$.  Under mean-field theory, the effective mass $m^*$ is equal to the atomic mass.
\index{Nambu-Goldstone bosons|)}

\subsection{Coherent atom optics with magnon excitations}
\label{sec:createmagnon}

Magnons can be written onto a spinor Bose gas by realizing a spatially dependent rotation of the gas magnetization.  \index{magnons!optical imprinting} Such rotation can be effected using the vector ac Stark shift of an off-resonant light field \cite{cct72}, which imposes an effective magnetic field oriented along the helicity of the light field and is proportional in strength to a product of the light ellipticity and intensity.

To see how this light-induced energy shift creates magnon excitations, consider the effect of light with $\sigma^+$ circular polarization along the $\mathbf{x}$ axis, with an intensity pattern of $I(\mathbf{r})$.  The vector ac Stark shift \index{vector ac Stark shift} creates an effective inhomogeneity in the magnetic field $\Delta \mathbf{B}(\mathbf{r}) \propto I(\mathbf{r}) \mathbf{x}$.  Suppose a fully polarized ferromagnetic condensate is prepared with its magnetization uniform, and oriented transversely to a magnetic field along $\mathbf{x}$.  The vector ac Stark shift now causes the condensate magnetization to precess by an additional angle $\Delta \phi \propto I(\mathbf{r})$ in the transverse plane.  This inhomogeneous spin texture represents a coherent wavepacket of magnons (realized in Ref.\ \refcite{veng07mag}).  Similarly, consider that the ferromagnetic condensate is prepared with uniform longitudinal polarization along a magnetic field oriented perpendicular to $\mathbf{x}$.  If the intensity or polarization of the incident light is now modulated at the Larmor precession frequency, the atoms will undergo Rabi oscillations due to the applied light field, again creating a coherent magnon excitation.

The magnons produced in this way were characterized by our ASSISI method, \index{imaging methods!absorptive spin-sensitive in-situ imaging} as shown in Figs.\ \ref{fig:magnonimage}.  To interpret these images, we recall that a rotation by the angle $\theta$ about the $\mathbf{n}_\phi$ axis transverse to the magnetization of a ferromagnetic spinor transforms the condensate wavefunction, to lowest order in $\theta$, as
\begin{equation}
\psi = \sqrt{n} \left( \begin{array}{c} 1 \\ 0 \\ 0 \end{array} \right) \rightarrow \sqrt{n} \left( \begin{array}{c} 1 \\ e^{i \phi} \theta / \sqrt{2} \\ 0 \end{array} \right).
\end{equation}
Therefore, to lowest order, a magnon excitation atop a condensate in the  $|m_F = +1\rangle$  internal state is just a single atom in the $|m_F = 0\rangle$ state.  Therefore, an image of the $m_F = 0$ atomic population is effectively an image of the (small) magnon density.

\begin{figure}[t]
\begin{center}
\includegraphics[width=5in]{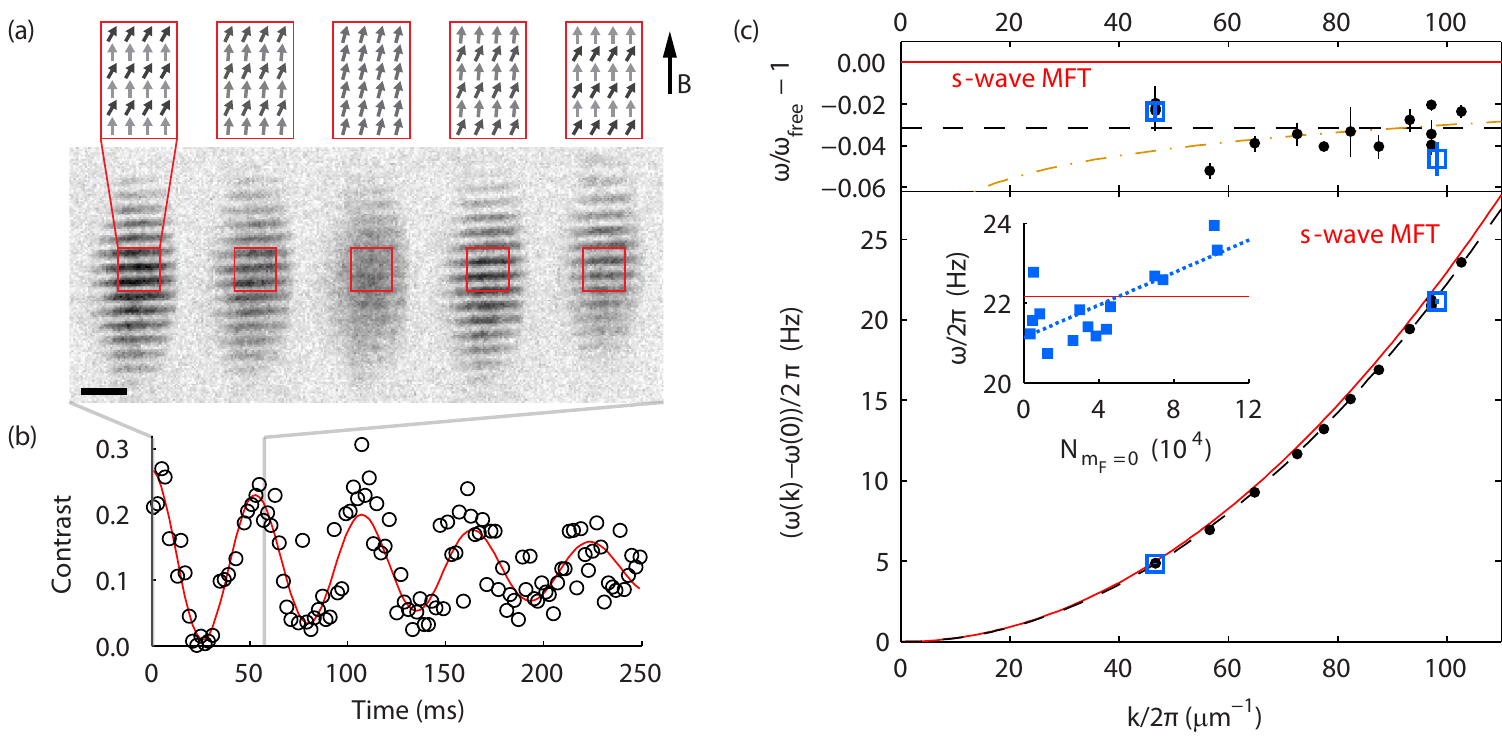}
\end{center}
\caption{In our magnon interferometer, a standing wave of magnons is imprinted onto the condensate, being composed of magnon waves at three distinct wavevectors.  Free propagation of the magnon waves causes the contrast of the spatial modulation of the magnon density at a wavevector $\mathbf{q}$ oscillates sinusoidally at frequency $2 (E(q) - E(0)) / \hbar$. (a) Images of the $m_F = 0$ population during this free propagation shows high initial contrast, diminished contrast after half a temporal cycle, and the high contrast obtained after a full temporal cycle (with fringes $\pi$ out of phase with the initial pattern).  (b) The contrast shows clear temporal oscillations.  (c) From these oscillations, we obtain the magnon recoil energy, shown in closed black circles.  The recoil frequency is corrected for a frequency shift with the number of magnons produced in the interferometer, with the measured recoil energy being the value obtained by interpolation to zero density.  Data are compared to the theoretical prediction of mean field theory (red line) that the magnon recoil energy should equal the free-atom recoil energy.  Our data show the magnon mass to be slightly larger than the bare atomic mass.  Figure reproduced from Ref.\ \refcite{mart14magnon}.}
\label{fig:contrastdispersion}
\end{figure}

Consider that one prepares a spatial grating of magnons using the vector ac Stark shift of light beams intersecting with wavevector difference $\mathbf{q}$.  One thereby prepares a coherent magnon wave with the initial wavefunction \index{magnons!contrast interferometer|(}
\begin{equation}
\psi_M(t=0) \propto 1 + \eta \cos(\mathbf{q} \cdot \mathbf{r}) = 1 + \frac{\eta}{2} \left(e^{i \mathbf{q} \cdot \mathbf{r}} + e^{-i \mathbf{q} \cdot \mathbf{r}}\right)
\label{eq:psi0}
\end{equation}
where we have assumed the ferromagnetic condensate to be stationary and uniform.  The factor $\eta$ accounts for the visibility of the light interference pattern.  This density-modulated magnon wave is a superposition of magnons at three wavevectors: $\mathbf{k} \in \{0, \mathbf{q}, -\mathbf{q}\}$.  After evolving for a time $t$, the energy difference between stationary and moving magnons causes the magnon density distribution to become
\begin{equation}
|\psi_M(t)|^2 \propto 1 + \eta^2 \cos^2(\mathbf{q} \cdot \mathbf{r}) + 2 \eta \, \cos\left(\frac{E(q) - E(0)}{\hbar} t\right) \cos(\mathbf{q} \cdot \mathbf{r})
\end{equation}
We observe the amplitude of the density modulation at wavevector $\mathbf{q}$ is temporally modulated at the Bohr frequency $(E(q) - E(0))/\hbar$, providing an interferometric measurement of the magnon dispersion relation\index{magnons!dispersion relation}.  Such modulations are shown in Fig.\ \ref{fig:contrastdispersion}.  By Galilean invariance, the contrast modulation frequency is Doppler free.  Contrast interferometry was demonstrated in a scalar Bose-Einstein condensate, where the signal was strongly sensitive to density-dependent shifts of the Bragg resonance frequency \cite{gupt02inter}.

\renewcommand{\baselinestretch}{1}
\renewcommand{\baselinestretch}{1.5}

\index{magnons!contrast interferometer|)}

From a materials science perspective, a precise measurement of the magnon dispersion relation is a precise test of quantitative many-body theories that describe the interacting ferromagnetic Bose gas.  A recent theoretical calculation \cite{phuc13} that goes beyond mean-field theory to consider the effects of phonon-magnon interactions predicts an increase in the magnon mass above the bare atomic mass by a fractional amount of about $3 \times 10^{-3}$.  In our experiment, we find the magnon mass to be heavier than the atomic mass by a fractional amount of $3 \times 10^{-2}$\index{magnons!effective mass}.  The origin of this discrepancy is unknown, though one should note that the theoretical calculations were performed for a uniform zero-temperature gas with no magnetic dipole interactions\index{spinor Bose-Einstein gas!magnetic dipolar interactions}, whereas the samples used in our experiment were inhomogeneous, finite-sized, at non-zero temperature, and possessed discernable magnetic dipolar interactions (as were also seen in the experiment).  Given the precision of the magnon contrast interferometer and the flexibilty of the experimental setup, it is clear there will be scientific payoffs in continuing this investigation: probing gases of variable size and geometry, isolating the effects of dipolar interactions, varying the temperature in a controlled manner, probing different initial states, and so on.

From an atomic physics perspective, the remarkable observation is how close is the magnon mass to the atomic mass: In a gas with a density on the order of $10^{14} \, \mbox{cm}^{-3}$, interactions appear to shift the magnon recoil frequency only at the precent level, with the absolute value of the shift being less than a Hertz.  In contrast, the density-induced shifts in the Bragg resonance frequency would be orders of magnitude larger.

Indeed, the properties of ferromagnetic magnons have enticing implications for atom interferometry.  Their gapless nature implies that magnons are created without a density-dependent interaction energy shift. With the magnon mass being just about equal to the atomic mass, magnon excitations propagate just about like free particles.  Thus, an interferometer based on the free propagation of magnons within a ferromagnetic spinor Bose-Einstein condensate at equilibrium (equal chemical potential across the condensate) is equivalent to an atom interferometer using free particles at zero gravity and in vacuum.
\index{magnons|)}

\section{Spin-mixing instability in a spatially multi-mode system}
\label{sec:spinmixing}

\index{spinor Bose-Einstein gas!spin mixing instability|(}\index{quantum quench|(}
The Section above illustrates how the study of atomic materials can complement that of solid-state materials by allowing and justifying precise quantitative tests of theory.  Another advantage offered to the study of materials science is the study of materials far from equilibrium, which is possible in atomic gases owing to the vast separation of timescales between the equilibration time (from just below a millisecond to much longer) and the shortest time (e.g.\ microseconds) in which the system Hamiltonian can be varied or the atomic internal state can be modified.

One scenario considered for such non-equilibrium dynamics is the response of an initially equilibrated many-body system to a rapid change of the system Hamiltonian.  Similar to quenching a hot iron bar by submersing it in water, here we consider a quench of a quantum system to a condition that favors a different equilibrium state.  The many-body system is thus uniformly in a non-equilibrium state.  The deviations from equilibrium can propagate spatially across the system, allowing correlations and entanglement to span a distance that grows with time.  The deviations can also serve to nucleate instabilities \cite{lama12chapter}.

Several works by my group have examined the quench of a degenerate spinor Bose-Einstein gas across a symmetry breaking phase transition.  In particular, we consider the $^{87}$Rb $F=1$ spinor gas for which the spin-dependent contact interactions favor ferromagnetism, as discussed above.  Atop these interactions, we add a single-particle energy term, a quadratic Zeeman energy $q F_z^2$ along an experimentally selected axis $\mathbf{z}$.  For large positive $q$, this term favors a condensate fully polarized in the $|m_F = 0\rangle$ state quantized along the $\mathbf{z}$ axis.  The Hamiltonian of the system is now symmetric under rotations about $\mathbf{z}$ and also under uniform gauge transformations (multiplication of the system wavefunction by a uniform complex phase).  For large positive $q$, the non-magnetized equilibrium state breaks just the gauge symmetry, in establishing a complex, scalar superfluid order parameter.  For small positive $q$, the ferromagnetic state is magnetized transverse to the $\mathbf{z}$ axis, and thus breaks the axisymmetry of the Hamiltonian as well as gauge symmetry.   These states of different broken symmetry are separated by a phase transition, which can be traversed by changing $q$.  An additional change of symmetry occurs for $q<0$, where the favored longitudinally magnetized state breaks the discrete $\mathds{Z}_2$ time-reversal symmetry but retains axisymmetry.

Let us focus on a quench from the non-magnetized, axisymmetric phase at large positive $q$ to the symmetry broken ferromagnetic phase at small positive $q$.  How does a system quenched through this transition dynamically break the previously unbroken axisymmetry?

\subsection{The Kibble-Zurek picture of inhomogeneous symmetry breaking}

\index{Kibble-Zurek mechanism|(}In the study of materials one encounters many cases of systems that undergo symmetry-breaking transitions, typically after being rapidly cooled (conventionally quenched). Considering systems as conventional as ferromagnets or supercooled water, we observe that symmetry breaking occurs locally.  Small regions of the material undergo the phase transition independently, and these regions then serve to nucleate larger regions of commonly broken symmetry.  Where these regions abut, we observe defects -- domain walls in magnets or crystal defects in ice -- that signify a conflict between the choices of broken symmetry in different regions, and that heal only very slowly as these conflicts are resolved.\index{coarsening}

The same phenomena are understood to occur in quantum materials undergoing quantum phase transitions.  The picture often invoked to describe the symmetry breaking dynamics is called the ``quantum Kibble-Zurek mechanism,''  \cite{dzia05,polk05univ,zure05qpt}, following earlier considerations for classical thermal phase transitions \cite{kibb76,zure85cosmo}.  This picture divides the quench dynamics into three periods in time.  In the first, a quantum system quenched across a phase transition will form regions of commonly broken symmetry with a size that shrinks as the quench rate is increased, with the exact relation between size and rate determined by the character of the dispersion relation for excitations near the phase transition point.  Second, the jumbled symmetry breaking will result in a tangle of defects \index{topological defects} of various types that depend on the geometry of the order parameter manifold.  Third, these defects will slowly coarsen and heal over long times as the system becomes ordered at longer length scales.

So far, the predictions for the second and third periods of evolution have been examined in spinor Bose-Einstein condensates that are sufficiently large in either one or two dimensions so that it makes sense to invoke the Kibble-Zurek mechanism to describe their evolution.  The spatially inhomogeneous symmetry breaking following a very rapid (essentially instantaneous) quench across the aforementioned paramagnet-to-ferromagnet transition is shown in Fig.\ \ref{fig:quench} \cite{sadl06symm}.  Regions of transverse magnetization appear some tens of milliseconds after the quench.  Regions of differently oriented magnetization are joined either by extended regions in which the magnetization gradually varies from one orientation to another, or by sharp lines of apparently unmagnetized gas (although it is possible the gas within these lines remains fully magnetized, with the magnetization varying over a length scale below our imaging resolution).  The dynamics of symmetry breaking also give rise to topological defects\index{topological defects}: polar-core spin vortices that, in fact, turn out to be the only topologically distinct vortex for the $F=1$ ferromagnetic superfluid.  The quantum quench in a polar $F=1$ spinor condensate has also been studied \cite{book11}.

\index{coarsening|(}As for the third period of the Kibble-Zurek chronology, several elements of coarsening dynamics have been examined.  Early studies \cite{mies99meta,stam99tunprl} shed light on the growth of spin domains in one-dimensional spinor Bose-Einstein condensates through two distinct mechanisms: Just as the normal and superfluid components of superfluid helium show distinct transport characteristics, here the non-condensed component of the gas allows for coarsening through thermal activation and re-equilibration, while the condensed component of the gas allows for domain growth through coherent quantum tunneling across domain walls.

My group examined the process of coarsening in two dimensional systems.  We examined the evolution of an initially hot and unmagnetized gas after it was cooled across the Bose-Einstein condensation phase transition and allowed to equilibrate for several seconds.  Regions of magnetized gas were observed after crossing the transition\footnote{As far as we could tell, the condensation of the gas and the formation of magnetized spin domains occurred simultaneously.  However, according to our experimental protocol, for each gas sample we were able to measure either the presence of a condensate (in time-of-flight images) or the presence of transverse magnetization (in in-situ images).  It would be interesting to perform both probes in a single experimental realization so that one can resolve better whether Bose-Einstein condensation occurs temporally before the formation of magnetized domains.  I suspect that it does.}.  In contrast with the $\sim$10 ms timescale for the initial formation of spin domains, here we observe a seconds-long timescale for domain coarsening.  One disappointing conclusion of this study is that the equilibration timescales for the $^{87}$Rb $F=1$ spinor Bose-Einstein gas are so long as to be experimentally inaccessible for spatially extended gases.
\index{coarsening|)}\index{Kibble-Zurek mechanism|)}

\subsection{Symmetry breaking through parametric amplification of spin fluctuations}

Let us return to the question of how an isolated system spontaneously breaks symmetry.  If the initial state of the system is symmetric, and the Hamiltonian of the system conserves that symmetry at all times in the evolution, then the final state of the system is perforce symmetric.  In that case, it stands to reason that the states we observe at late times after the quench are indeed symmetric superpositions of states of macroscopically broken symmetry.

Quantum optics offers a candidate for such a superposition state: squeezed vacuum\index{squeezed states}.  Applying a phase-sensitive parametric amplifier to the vacuum state causes one quadrature of the electromagnetic field to be amplified.  Both the vacuum state and the squeezed vacuum state are symmetric, in that quadrature measurements are equally likely to give positive or negative values.  However, measurements on the squeezed state are likely to yield either a large positive or large negative value, deviating from the vacuum levels by potentially macroscopic amounts.

Perhaps, then, the dynamics of a many-body quantum system following a quantum quench may be described as the parametric amplification of initially symmetric and microscopic vacuum fluctuations.  Questions arise:  What is being amplified?  What is being squeezed?  Is the amplifier quantum-limited?  What is the gain of this amplifier?  If this amplifier is applied to a spatially extended condensate, the amplifier might be expected to act simultaneously but separately on fluctuations in different portions of the condensate.  If so, how are different spatial patterns of fluctuations affected by the amplifier?  In other words, what is the spatial gain spectrum of this amplifier?

Answers have been provided to several of these questions.  It is understood that the spin fluctuations in this case -- amplification through the spin-mixing instability of a gas initially in the $|m_F = 0\rangle$ quantum state -- exist in two independent planes, representing two different polarizations of magnon excitations of the initial state \cite{lesl09amp,sau10njp}\index{parametric amplifier of spin fluctuations}.  Each plane is spanned by two observables: one component of the tranverse spin and one component of the spin-quadrupole tensor, serving as the independent and non-commuting quadratures equivalent to the electric field quadratures of light or the position and momentum quadratures of a mass on a spring.  The transverse spin quadrature can be measured directly using circular birefringence dispersive imaging \index{imaging methods!spin-sensitive dispersive} \cite{higb05larmor,sadl06symm,lesl09amp}.  The spin-quadrupole moment can be measured by first rotating it onto the transverse-spin quadrature by briefly applying a strong quadratic Zeeman shift, as has been demonstrated in single-mode experiments probing this instability \cite{haml12squeezing}.

\begin{figure}[t]
\begin{center}
\includegraphics[width=3in]{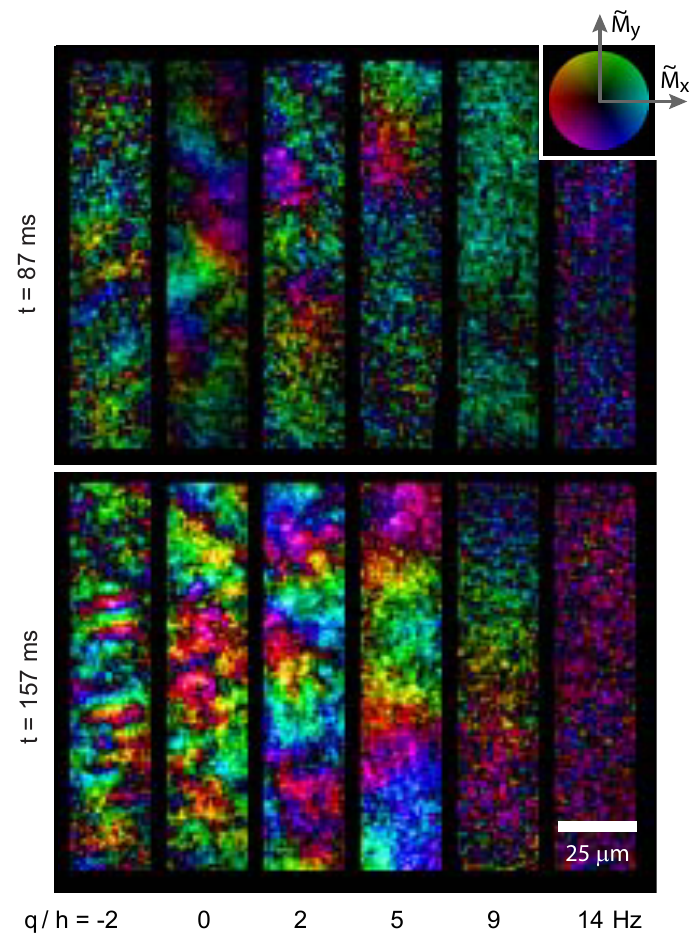}
\end{center}
\caption{An electronic amplifier has a spectrum, describing the fact that gain at particular temporal frequencies is stronger than at other frequencies.  Similarly, in a spatially extended spinor Bose-Einstein condensate, quenching to conditions that favor symmetry breaking into a ferromagnetic states causes the condensate to act as a spin-fluctuation amplifier that has a gain that depends on the spatial mode structure of the fluctuation.  In particular, quenching the system to lower values of the quadratic Zeeman shift (images further to the left) causes spin fluctuations with smaller wavelength to be amplified more strongly.  This spatial gain spectrum is visible in images taken at variable times after the quench.  The initial spin noise in the spinor condensate, which we presume to be present at equal strength at all wavevectors (spatially white) is amplified by a wavelength-dependent spin amplifier.  The macroscopic magnetization spectrum produced by this quench shows the spatial lengths most favored by the amplifier.  Each image here is a map of the tranverse magnetization in a single condensate, at one of two different times after the quench (indicated on the left) at several different values of the quadratic Zeeman shift (indicated on the bottom).  Figure taken from Ref.\ \refcite{lesl09amp}.}
\label{fig:amplifier}
\end{figure}

For spatially extended systems, one can characterize the gain of this amplifier in momentum space, just as one characterizes an amplifier that acts on a continuous signal by its gain spectrum in frequency space.  Through a linear stability analysis of the Gross-Pitaevskii equation that describes the coherent evolution of an initially uniform spinor Bose-Einstein condensate, one finds the magnon dispersion relation \index{magnons!dispersion relation}\index{magnons!instability}for the $|m_F = 0\rangle$ initial state, given as $E_s^2 = (\epsilon_k  + q) (\epsilon_k + q - |c_1^{(1)}| n)$ where $\epsilon_k = \hbar^2 k^2 / 2 m$ is the free particle kinetic energy at wavenumber $k$ and $n$ is the condensate density.  For $q < q_0 = 2 |c_1^{(1)}|n$, there is a range of wavenumber where $E_s^2<0$ indicating a regime of parametric instability with a temporal gain given by $\sqrt{|E_s^2|}/\hbar$.  The colored spectrum of this amplifier can be interpreted as defining particular localized spatial modes that are amplified with highest gain \cite{lama07quench,sau10njp}.  Rough features of this gain spectrum, and how it tunes with the quadratic Zeeman shift, were observed in Ref.\ \refcite{lesl09amp} (Fig.\ \ref{fig:amplifier}).

\renewcommand{\baselinestretch}{1}
\renewcommand{\baselinestretch}{1.5}

The noise characteristics of this parametric amplifier have been examined in single-mode spinor systems, where it has been shown to yield highly spin squeezed states \cite{haml12squeezing}. My group attempted to characterize the noise limits of this parametric amplification in a spatially extended spinor gas, using our high-resolution imaging to quantify the amplifier output after a variable time of amplification \cite{lesl09amp}.  Unfortunately, our uncertainty in the temporal gain of the amplifier, stemming from uncertainty in the scattering lengths of $^{87}$Rb, did not permit us to place tight constraints on the amplifier noise.
\index{spinor Bose-Einstein gas!spin mixing instability|)}\index{quantum quench|)}

\section{Spatially resolved magnetometry with a quantum fluid sensor}
\label{sec:magnetometry}

\index{magnetometry!spatially resolved|(}\index{magnetometry!comparison of methods|(}Finally, let me consider the use of cold atomic gases for spatially resolved magnetometry to study materials.  At present, several technologies are used for spatially resolved magnetometry at micron length scales, including superconducting quantum interference devices (SQUIDs), scanning Hall probe microscopes, magnetic force microscopes and magneto-optical imaging techniques \cite{bend99magnetic}.  Among these, the technologies offering the highest field sensitivity for single-pixel measurements are the SQUID microscope and the scanning Hall probe microscope.  SQUID sensors excel at measuring high-frequency signals, e.g.\ in the GHz range, where quantum-limited performance has been achieved \cite{muck01squid}, equivalent to roughly $S = 1 \, \mbox{pT}^2/\mbox{Hz}$ sensitivity in a $100 \, \mu\mbox{m}^2$ measurement area.  For low frequency signals, the performance of SQUID microscopes is compromised by $1/f$ flux noise, achieving $S \simeq (40 \, \mbox{pT})^2/\mbox{Hz}$ in a $100 \, \mu\mbox{m}^2$ measurement area \cite{kirt95mag,lee96mag}.  The scanning Hall probe microscope achieves a worse sensitivity of $S \simeq (100 \, \mbox{nT})^2/\mbox{Hz}$, but its high spatial resolution (optimal around 200 nm) makes it competitive with SQUID sensors at length scales at or just below one micron.

Both these technologies require the sensor to be scanned sequentially across the field of view of the microscope, reducing the amount of measurement time available at each resolved pixel.  Thus, magnetic field images with many resolved pixels are acquired with correspondingly worse measurement sensitivity.  This disadvantage is circumvented by the use of magneto-optical imaging, which relies on the linear Faraday effect (birefringence) of solid-state materials.  Magneto-optical imaging allows for the simultaneous acquisition of a magnetic-field image across a wide field of view, with spatial resolution around 1 $\mu$m matching the wavelength of light used for imaging.  The sensitivity of this technique in present implementations is lower, at single pixels, than the SQUID and scanning-Hall probes described above.  However, this worsening of sensitivity is made up by the parallelism of measuring across a wide field of view.

These magnetic microscopy methods are used in an extensive range of applications.  Scientific applications include imaging vortex structures in superconductors \cite{bend99magnetic,guik08} and resolving their temporal dynamics \cite{plou00dynamics} and probing for spontaneous magnetization in $p$-wave superconductors \cite{kirt07}.  A major technological application is non-destructive evaluation of metallic objects and microfabricated devices by means of detecting eddy currents or bound currents in stressed ferromagnetic materials, susceptibility imaging, and detailed mapping of current flow \cite{kirt99review}.

\index{magnetometry!comparison of methods|)}

\index{spinor Bose-Einstein gas!magnetometry|(}A cold-atom magnetometer is compatible with many of the applications discussed above, and yet achieves a single-pixel sensitivity that is orders of magnitude better than the SQUID and scanning Hall probe sensors while maintaining the advantage of detecting over a wide field of view in a single measurement.  The possibility of using spinor Bose-Einstein gases for high precision magnetometry was established by prior work of my group \cite{veng07mag}.  In that work, a dense, spin-polarized Bose-Einstein condensed gas of $^{87}$Rb was confined in a near-planar geometry in a far-off-resonant optical dipole trap.  A measurement was begun by tipping the spin polarization to lie transverse to an applied magnetic field, with a spatially uniform initial orientation.  The gas was then allowed to undergo Larmor precession, at an angular frequency determined by the local magnetic field magnitude $B(\boldsymbol{\rho})$ as $\omega_L(\boldsymbol{\rho})=\gamma B(\boldsymbol{\rho})$ where $\gamma = g_F \mu_B/\hbar$ is the atomic gyromagnetic ratio\footnote{Conventionally, $\gamma$ would be defined with a minus sign.  We use $\gamma$ here to relate the Larmor frequency directly to the magnetic field.} and $\boldsymbol{\rho}$ indicates position in the imaged 2D plane. After a measurement time $\tau$, the Larmor precession phase $\phi (\boldsymbol{\rho})$ was measured by high-resolution magnetization-sensitive imaging.

Let me highlight several features of that work.  First, the cold atom gas represents an extended sensing field, providing a simultaneous spatial (2D) map of the field magnitude.  Compared with a scanning single sensor, such as a scanning SQUID microscope, this simultaneous mapping rejects long-spatial-wavelength temporal field fluctuations, which appear as a spatially uniform shift of the Larmor phase $\bar{\phi}$, to reveal static short-wavelength spatial inhomogeneities, which are measured as $B(\boldsymbol{\rho})-\bar{B} = \frac{1}{\gamma \tau}( \phi(\boldsymbol{\rho}) - \bar{\phi})$.  Such common-mode rejection bestows the practical advantage of achieving superior magnetic field sensitivity even in the absence of any magnetic shielding.

Second, the sensitivity of this proof-of-principle experiment already approached the standard quantum limit for magnetic sensing. \index{magnetometry!standard quantum limit}This limit derives from atomic-shot-noise limit to phase measurements, $(\Delta \phi)^2 = N^{-1}$, where the number of atoms in a resolved pixel of area $A$ is given as $N = n_{2D} A$ with $n_{2D}$ being the areal atom number density.  The measurement sensitivity is then given as
\begin{equation}
S(A) = \frac{1}{\gamma^2} \frac{1}{A} \frac{1}{n_{2D} \tau D}\label{eq:S}
\end{equation}
The sensitivity, with units $[\mbox{T}^2/\mbox{Hz}]$, scales inversely with the areal resolution, and improves with higher atom column density ($n_{2D}$), single-measurement time $\tau$, or duty cycle $D \leq 1$ which is the ratio of the single-measurement time to the cycle time between measurements.

\begin{figure}[t]
\begin{center}
\epsfig{file = 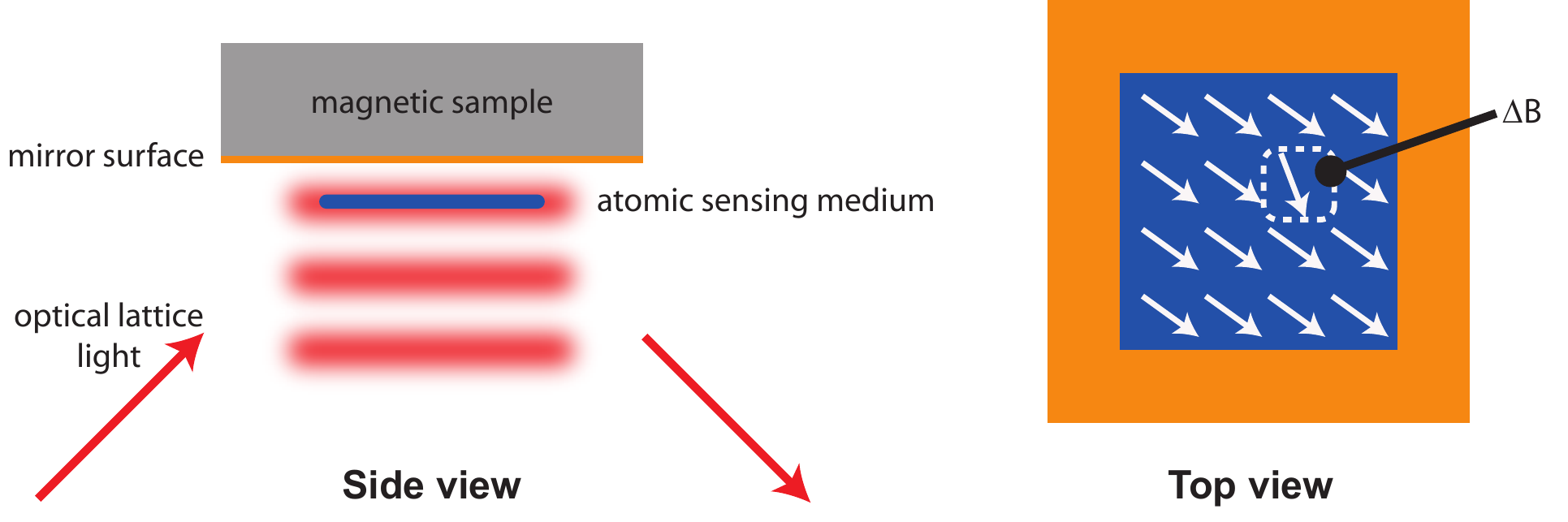, height = 1.5in} \caption{\small Magnetometery scheme.  Side view: A mirror surface on a magnetic sample is used to produce a vertical optical lattice from laser light reflected off the mirror surface.  The incident angle of the laser light controls the lattice spacing.  Top view: Following Larmor precession for the single-shot measurement time, the orientation of the transverse magnetization or nematicity is imaged at high spatial resolution across the atomic sample.  Local inhomogeneities of the magnetic field are visible in the inhomogeneous orientation of the atomic spin.} \label{fig:scheme}
\end{center}
\end{figure}

Now, consider implementing this measurement with a layer of ultracold atoms trapped at micron distances from a mirror surface, where they serve as a sensor for magnetic materials below, on, or above the surface (Fig.\ \ref{fig:scheme}).  For example, the atoms might be trapped within an optical lattice trap, formed by light reflecting at shallow angles off the surface of the material being probed or off a thin ``cover slip'' placed above the material.  The quality of the optical trap is not critical: distortions of the potential by scattered light will distort the atomic density distribution, but will not affect the Larmor precession of the trapped atoms.

Once placed within the surface-supported optical lattice trap, the spin-polarized atoms can serve as a sensor of steady (dc) or alternating (ac) magnetic fields.  For this, the atomic spin is prepared in a spatially uniform, transversely polarized, initial state, allowed to evolve for a measurement time $\tau$ in the presence of the field produced by the sample, an applied bias field, and rf pulses.  Last, the atoms are probed using state-sensitive imaging to reveal the phase of Larmor precession with high spatial resolution, and the imaging data processed to derive the target magnetic field.

\renewcommand{\baselinestretch}{1}
\renewcommand{\baselinestretch}{1.5}

The quantum-limited sensitivity to be expected from this device is derived from Eq.\ \ref{eq:S}.  Consider that atoms are trapped in the optical-lattice site closest to the reflective material surface, at a distance $d = (\lambda /4) \csc \theta$ from the mirror surface, where $\lambda$ is the wavelength of the light forming the lattice and $\theta$ is its angle of incidence.  With $\lambda$ on the order of 1 $\mu$m, $d \geq 250$ nm.  Let us assume a three-dimensional density of $n_{3D} = 3 \times 10^{14} \, \mbox{cm}^{-3}$, a thickness of $d/4$ for the lattice-confined atomic layer, and single-shot measurement time of $\tau = 3 \, \mbox{s}$.  One then obtains
\begin{equation}
S(A) \simeq \left(1.5 \, \frac{\mbox{pT}^2}{\mbox{Hz}} \right) \times \left(\frac{\mu\mbox{m}^2}{A} \frac{\mu\mbox{m}}{d} \frac{1}{D} \right) \label{eq:Swithnum}
\end{equation}
This measurement sensitivity is far superior to existing dc SQUID magnetometers (limited at $\sim(10\, \mbox{nT})^2/\mbox{Hz}$) which are regarded presently as the most powerful magnetometers at the micron length scale.  Additional benefits of such a cold-atom magnetometer include the following:

\emph{Absolute calibration:}  The magnetic sensitivity for this cold-atom device is based directly on atomic properties.  In contrast, microfabricated SQUID magnetometers are subject to calibration errors and drifts.  Magnetometers based on solid-state spin impurities are strongly influenced by their inhomogeneous environments, so that each impurity must be individually diagnosed.

\emph{Simultaneous measurement across a wide spatial region:}  The cold-atom magnetometer envisioned here excels at differential measurements of the magnetic field obtained in parallel in many resolved pixels of the sensor surface.  This differential sensitivity provides three important advantages.  First, long-wavelength magnetic disturbances result in common-mode shifts of the Larmor phase and are thus rejected.  Thus, the influence of any source of stray magnetic fields further than 100's of microns from the sensor is essentially eliminated, obviating the need for strict controls or shielding of the magnetic environment.  Second, this feature obviates the need to register precisely the positions of a magnetic object and the sensor; rather, the magnetic object is simultaneously located and measured by the cold-atom sensor.  Third, making simultaneous measurements over a total sensor area $A_{tot}$ at many resolved pixels, each of area $A$, provides an additional improvement in sensitivity, reducing $S$ in Eq.\ \ref{eq:S} by a factor $A/A_{tot}$, compared to a single scanning sensor.

\emph{Operation at high temperature:}  Atoms can be maintained at ultralow temperature even within micron distances of high-temperature objects \cite{obre07casimir}.  Thus, a cold-atom magnetometer can measure magnetic objects over a wide temperature range, e.g.\ for the study of magnetic phase transitions.

\emph{Operation at high magnetic field:}  While high fields disturb the operation of superconductor-based magnetic sensors, the immunity of our cold-atom magnetometer to uniform bias fields allows high sensitivity to be achieved even in a strong uniform applied magnetic field.  This allows for sensitive detection of magnetic susceptibility.

Let us quantify the ideal sensitivity of such a cold-atom sensor for several applications.

\emph{Detecting currents:}  \index{magnetometry!detecting currents|(}First, consider the task of detecting currents on or within the sample, an essential component in many applications of magnetometry for non-destructive evaluation of materials and devices, including detection of conducting materials by the generation of eddy currents, the identification of disturbances in current flow through conductors, the assessment of material defects by the hysteretic magnetization of stress-modulated materials, and the characterization of failures in integrated circuits \cite{jenk97nde,kirt99review}.  A line current $I$ running parallel to the mirror surface at a depth of $r$ produces a magnetic field of magnitude $B = \mu_0 I / (2 \pi r)$.  We compare this field to the single-shot atom-noise-limited measurement sensitivity $S(A)_{single} = \frac{1}{\gamma^2} \frac{1}{A n_{2D} \tau^2}$.  We consider the task of resolving features of the field produced by this wire (or many wires) with areal spatial resolution of $A = d^2$ set by the distance of the atoms from the mirror surface.  As above, we take $n_{2D} = n_{3D} \times d/4$ and $n_{3D} = 3 \times 10^{14} \, \mbox{cm}^{-3}$, and single-shot measurement time $\tau = 3$s.  This gives a signal-to-noise ratio of unity for a current of
\begin{equation}
I = 4 \, \mbox{pA} \times \left(\frac{\mu\mbox{m}}{d}\right)^{3/2} \times \left( \frac{r}{\mu\mbox{m}}\right)
\end{equation} \index{magnetometry!detecting currents|)}

\emph{Detecting permeable materials:}   \index{magnetometry!detecting permeable materials|(}Second, consider the task of detecting a spherical structure of radius $R$, composed of material with magnetic susceptibility $\chi_m$, and buried within the substrate at a distance $r > R+d$ from the atomic sensor layer. Let us assume the surrounding material has susceptibility $\chi_m = 0$.  In an applied magnetic field $\mathbf{B}_0$, the spherical structure acquires a uniform magnetization of $\mathbf{M} \simeq (\chi_m/\mu_0) \mathbf{B}_0$.  Outside the sphere, the induced magnetization produces a field equivalent to a magnetic dipole of moment $\mathbf{m} = (4/3) \pi R^3 \mathbf{M}$, i.e.\ the field produced at the atomic gas above the mirror surface has magnitude $B \simeq (\mu_0/4 \pi) |m|/r^3$.  This induced field is nearly uniform over an area $A \sim r^2$ across the atomic gas; taking this as the resolving area of the magnetometer, we conclude a single-shot measurement will detect a susceptibility of order
\begin{equation}
\chi_m = 3 \times 10^{-8} \times  \left(\frac{\mu\mbox{m}}{d}\right)^{1/2} \times \left( \frac{r}{\mu\mbox{m}}\right)^2 \times \left(\frac{\mu\mbox{m}}{R}\right)^{3} \times \left( \frac{\mbox{G}}{B_0} \right)
\end{equation}
Thus, at just a few Gauss applied field, micron-scale structures with susceptibility below $10^{-8}$ should be detectible; this is well below the typical susceptibility of diamagnetic solids ($10^{-5}$) and equal to those of diamagnetic gases ($10^{-8}$). \index{magnetometry!detecting permeable materials|)}

\emph{Detecting surface spins:}  \index{magnetometry!detecting surface spins|(} Finally, we consider the task of locating and quantifying magnetic particles on the mirror surface, at distance $d$ from the atomic sensor.  Let these particles have a magnetic moment equal to $N_e \mu_B$ (i.e.\ $N_e$ commonly aligned electron spins).  Using the areal resolution $A=d^2$, the single-shot sensitivity is equivalent to
\begin{equation}
N_e = 0.9 \times \left(\frac{d}{\mu\mbox{m}}\right)^{3/2}
\end{equation}
Thus, even nanoscale ferromagnetic particles should be detectible. \index{magnetometry!detecting surface spins|)}

\index{spinor Bose-Einstein gas!magnetometry|)}\index{magnetometry!spatially resolved|)}

\section{Conclusion}

There remain many things to be learned about magnetic phenomena in quantum gases, and methods of direct imaging may allow us to investigate them.  One area that is clearly ripe for investigation is the detailed dynamics of spin textures, domains, domain walls, and vortices.  Our methods for imaging are now sufficiently refined that we should be able to investigate such dynamics by taking a set of images of a single gaseous sample as it evolves in time.  It will be interesting to track the motion of magnetic features so as to understand how they change and influence one another.  Such investigations will reveal the basic elements of superfluid hydrodynamics as it occurs in magnetic quantum gases, where the flow of mass and spin currents are interrelated.  It will also be valuable to measure precisely the temporal spin correlation function, which can be quantified by examining a pair of consecutive images taken of an evolving gas.  For systems in thermal equilibrium, this correlation function is related to the nature of spin excitations through the fluctuation-dissipation theorem.

Another open area of investigation is the effect of non-zero temperature on magnetic order and dynamics.  Much of the literature on spinor Bose-Einstein condensates considers their zero-temperature properties.  At non-zero temperature, we know that additional phenomena will arise, such as ``spin-locking'' \cite{mcgu03normal} and spin waves driven by the exchange interaction \cite{mcgu02}.  Both these phenomena have been observed with spatial resolution in pseudo-spin-1/2 gases, but detailed investigations in higher-spin spinor gases are lacking.  There are several types of experiments where very high quality data are being produced on lowest-temperature gases, such as the studies of quantum spin mixing dynamics and our recent precision measurement of the magnon dispersion relation.  Extending these studies carefully to higher temperature gases might help reveal essential differences stemming from the dynamics of non-condensed atoms.

Finally, we still await applications of spinor-gas magnetometry to the studies of magnetic materials.  Our measurement of the magnetic dipolar interactions in the rubidium spinor Bose gas, identified as a gap in the magnon excitation spectrum in Ref.\ \refcite{mart14magnon}, was an application of spinor-gas magnetometry to measure magnetic gases through the magnetic field they generate\index{spinor Bose-Einstein gas!magnetic dipolar interactions}.  However, as sketched in this Chapter, there are excellent prospects for studying solid-state materials by placing a spinor-gas sensor nearby.  These applications will be boosted by quantum effects such as spin-nematic squeezing and by improvements in spin-sensitive imaging.

\section*{Acknowledgements}

I thank the members of my research group who worked with me on spinor Bose-Einstein gases over the years.  In particular, I acknowledge the ``E4 team'' of G.\ Edward Marti, Ryan Olf, Andrew MacRae, Fang Fang, and Sean Lourette for inspiring a new set of ideas about how to improve spin-dependent imaging and how to characterize magnon excitations.  This work is supported by the NSF, NASA and by the AFOSR's MURI program on advanced quantum materials.

\bibliographystyle{ws-rv-van}

\printindex                         
\end{document}